\documentclass[onecolumn,amsmath,showpacs,amssymb,aps,nofootinbib,floatfix]{revtex4-1}
\usepackage{epsfig,amsmath,amssymb}
\usepackage{dcolumn,color}
\usepackage{bm}
\usepackage{graphicx}
\usepackage{subcaption}
\usepackage{float}
\usepackage{cleveref}

\newcommand{\beq}{\begin{equation}}
\newcommand{\eeq}{\end{equation}}

\newcommand{\ave}[1]{\left\langle #1 \right\rangle}

\begin{document}
\title{Causality of polarizeable dissipative fluids from Lagrangian hydrodynamics}
\author{David Montenegro$^1$, Giorgio Torrieri$^2$}
\affiliation{$^1$ BLTP,JINR,Dubna Russia\\$^2$ IFGW, Unicamp, Campinas, Brasil}
\begin{abstract} We perform a causality analysis on the dispersion relation of hydrodynamics with spin well as shear and bulk viscosity, including the relaxation times for all these quantities. We find that the interplay of the three relaxational scales, for shear and bulk viscosity as well as polarization, leads to non-trivial effects on the dispersion relation.   Unexpectedly, the presence of polarization leads to lower effective viscosity and a longer relaxation time, and the presence of viscosity leads to lower limits as well as upper ones on the group velocity and constraints relating polarization to viscosity relaxation times. We conclude with a qualitative discussion on how these results impact phenomenology, specifically the low effective viscosity in strongly interacting matter as well as shear-vorticity coupling.
\end{abstract}
\maketitle
\section{Introduction}
The applicability of relativistic hydrodynamics in heavy ion collisions has been a great phenomenological success \cite{snellings} and also spurred a lot of progress in the theoretical study of the subject \cite{rocha,disconzirev}.  Through this effort hydrodynamics has become a ``mature`` effective theory both in it`s mathematical consistency and in its quantitative ability to fit and predict soft experimental observables in heavy ion collisions.  While details of the expansion are still debated, it is clear that viscosity of the medium created in heavy ion collisions is low with respect to entropy density, hence hydrodynamics should be close to ideal, with sub-leading dissipative corrections.

In such a regime it is an elementary result that fluids are highly vortical \cite{rezzolla}. Furthermore, the thermodynamics of angular momentum can be used to connect vorticity to observable spin in hadrons \cite{review}.
Reasearch in this direction was also highly successful:
the discovery of vortical spin alignment of quark-gluon plasma has led to a fertile theoretical and experimental investigation of relativistic vortical matter \cite{review}.   Polarization probes of vorticity are becoming useful probes of phenomenology \cite{review,becshear,Schenke,jets1,jets2}
On the theory side, the central problem is to define the ideal hydrodynamic limit of matter where polarization is non-zero \cite{Koichi,polhydro1,polhydro2,causality,causality2,causality3,disconzirev,lagreview,bec1,flork1,flork2,gursoy,vortsus,shu,zubspin,lin1,lin2,chiarini}

A promising approach in this sense is Lagrangian hydrodynamics \cite{polhydro1,polhydro2,lagreview}, where the dynamics is dictated not by conserved currents and constitutive relations, but by a lagrangian\cite{lagorig}.   Because of this, the pseudogauge issue plaguing other approaches \cite{pseudo,pseudo2} is avoided in a way that can be easily generalized to higher spins \cite{ghosts} and a straight-forward link can be made between the thermodynamic free energy and the hydrodynamics \cite{lagreview}.

A significant achievement in this approach was proving, in a straight forward way \cite{causality} that when polarization is present dynamics becomes inherently dissipative, just from causality requirements.  This has been seen in other approaches \cite{causality2,causality3} and can be used to conjecture, with support from linear response \cite{lin1,lin2,kadanoff} and stability analysis \cite{koide,fogaca,Romatschke} that microscopic polarization is what stabilizes fluid dynamics as an effective theory against fluctuations carrying vorticity \cite{nicolis,rattazzi} (the resulting microscopic non-equilibrium dynamics in the form of shear-polarization coupling \cite{lin2} is similar to what was derived in other approaches \cite{shear1,shear2} can also be seen phenomenologically \cite{becshear}).

In this work, we therefore continue in this approach by combining insights from linear response \cite{lin1,lin2} and stability analysis \cite{koide,fogaca} to study the dispersion relation of linearized perturbations of dissipative hydrodynamics with spin, generalizing the results of \cite{causality} with an intrinsic shear and bulk viscosity.  

That this profoundly changes the results of \cite{causality} is not surprising:  Instead of one relaxation time-scale (spin-vorticity) as in \cite{causality}, we now have three, $\tau_Y$ regulating the relaxation of causality and vorticity and $\tau_{\eta,\zeta}$, the relaxation time-scales of the shear and bulk parts of the energy-momentum tensor.
As already found in \cite{lin2}, these three timescales combine in ways which resist a simple dependence. 

Indeed, dispersion relations exhibit a complicated interplay of these scales.
We shall also find that causality and stability lead to constraints, analogous to \cite{koide}, linking equilibrium and non-equilibrium coefficients characterizing both polarization and viscosity. 

In Section \eqref{background} we shall overview the basic results of lagrangian hydrodynamics and its application to fluids with spin.   In Section \eqref{disp0} we shall proceed to calculate the dispersion relations in the presence of additional dissipative forces.  Finally, in section \eqref{disp1} we shall study how all dissipative forces interplay together.  Section \eqref{disc} summarizes our results.
We note that a similar analysis was undertaken recently in \cite{chiarini}, but for Eulerian rather than Lagrangian hydrodynamics in the semi-classical approximation.
\section{Background\label{background}}
\subsection{Lagrangian hydrodynamics without and with polarization}
 We shall summarize the results of EFT (Effective Field Theory) applied to hydrodynamics \cite{polhydro1,polhydro2,lagorig,lagreview,nicolis,rattazzi}.   The basic degrees of freedom of lagrangian hydrodynamics without polarization are the Lagrangian coordinates $\phi_{I=1,2,3}$.  The dynamics is dictated by the symmetries of ideal hydrodynamics, namely the invariance under volume preserving diffeomorphisms (which polarization spontaneusly breaks \cite{polhydro1,lin2} in the local isotropy subgroup, as shown in table 2.1 of \cite{lin2}).  The Lagrangian as a result must depend only on the entropy density
The entropy density  
\begin{equation}\label{first}
b = \left( \det_{IJ} \left[ \partial_\mu \phi_I \partial^\mu \phi_J \right] \right)^{1/2} = \left( \det_{IJ} B^{IJ} \right)^{1/2}.
\end{equation} 
We can arbitrarily choose the "vacuum state" or equilibrium state in such a way that we choice the isotropic and homogeneous static configuration of the fluid $ \langle \phi^I \rangle = x^I$. Following the Poincaré invariance, it is extremely useful to introduce $K^\mu\partial_\mu\phi^I  = 0$. Respecting the symmetries of the field, we define the geometrical construction of the the vector as

\begin{equation}\label{K}
K^\mu  \equiv  \frac{1}{6} \epsilon^{\mu\alpha_1\alpha_2\alpha_3} \epsilon_{IJK} \partial_{\alpha_1} \phi^I\partial_{\alpha_2} \phi^J \partial_{\alpha_3} \phi^K.  
\end{equation}
It is straightforward to notice the antisymmetric propriety conserves this vector  
\begin{align}\label{Kconser}
\partial_\mu K^\mu=0.
\end{align}
The physical motivation is to build the Lorentz four-velocity $u^\mu \equiv K^\mu / b $ whose norm is $u^\mu u_\mu = 1$. Around a hydrostatic background, excitations can be expanded as
\begin{equation}
 \phi^I(t,x) = \left( x^i \delta^I_i + \pi^{I}(t,x) \right).
\end{equation}
\textcolor{black}{where the $\pi^{I}$ are massless Goldstone bosons since the static configuration conserves only internal translation and spatial symmetries as well as diagonal translation. Such expansion} gives rise to 3 degrees of freedom, which can be written as $\vec{\pi} = \vec{\pi}_T + \vec{\pi}_L$, with a dispersion relation\footnote{The notation \[\ \cdots \partial_\mu \pi\cdot \dot \pi =  \sum_I \cdots\partial_\mu \pi^I \dot \pi^I,  \qquad \dot \pi^2 = \sum_I \dot \pi^I \dot \pi^I, \qquad \delta^i_J \left( \partial_i \pi\cdot \partial \pi  \cdots \partial\pi^J \right) = [\partial \pi\cdot \partial \pi\cdots \partial \pi]. 
\] is used throughout this paper}
\begin{eqnarray}\label{omega^2_III}
 \vec{\nabla} \cdot \vec{\pi}_T &=& 0 \rightarrow \vec{\pi}_T = \vec{\nabla} \times \vec{\Omega}  \nonumber \\
 \vec{\nabla} \times \vec{\pi}_L &=& 0 \rightarrow \vec{\pi}_L = \vec{\nabla} \Phi 
\end{eqnarray}
It is convenient to introduce the Fourier transform 
\begin{equation}\label{Fouriertrans}
\left(
\begin{array}{c}
\Phi \\
\Omega\\
\end{array}
\right) =   \left(
\begin{array}{c}
\Phi_0 \\
\Omega_0
\end{array}
\right) \exp \left[  i \left( w_{L,T,Y} t -\vec{k}.\vec{x}  \right)\right]
\end{equation}
Now we can find the dispersion relation for longitudinal excitation $\pi^I_L$ and transverse ones $\pi^I_T$

\begin{equation}\label{ftdr}
 \pi^I_L = -i \phi^0 \partial_I \, e^{ i ( {\bf k}\cdot {\bf x}  -i\omega_L t ) } = k^I  \phi^0 \, , \qquad \pi_T^I = -i \epsilon^{IJK} \partial_J \Omega^0_K \, e^{i ( {\bf k}\cdot {\bf x} -i\omega_T t ) } = ({\bf k}\times {\boldsymbol \Omega^0})^I e^{ i ( {\bf k}\cdot {\bf x}  -i\omega_T t )  }.
\end{equation}
We are going to use later the full second order expression for the four-velocity
\[  u_\mu \simeq \delta^\mu_0 \left( 1+ \frac{1}{2}\dot \pi^2 \right) + \delta^\mu_I \left( \vphantom{\frac{}{}} -\dot \pi^I + \dot \pi\cdot\partial\pi^I  \right)  \]
and $\Delta_{\mu\nu} = g_{\mu\nu} - u_\mu u_\nu = B_{IJ} \partial_\mu \phi^I   \partial_\mu \phi^J$ whose properties  $ \Delta_{\mu\nu} u^\mu = \Delta_{\mu\nu} u^\nu =  0.  $ The first order expansion of this projector $\Delta_{\mu\nu}  \simeq ( g_{\mu\nu} - g_{\mu 0}\, g_{\nu 0}) + \left( g_{\mu 0} \, g_{\nu J} + g_{\mu J} \, g_{\alpha 0} \right) \dot \pi^J.$ We are only going to add the shear and bulk viscosity to a polarizable fluid. Our system follows the linear sum of the Lagrangians \cite{lagorig,lagreview}
\begin{align} 
& {\cal L} =  {\cal L}_{free} + {\cal L}_{Bulk} + {\cal L}_{NS} + {\cal L}_{Pol},  \\
& {\cal L}_{free} = F(b) \\
& {\cal L}_{Bulk} = \sum_{i,j,k} h_{ijk}( b^2 ) K^{\mu I} K^{ \nu J}  \partial_\mu K^K_\nu  \label{lagbulk} \\
& {\cal L}_{NS} = \sum_{i,j,k} z_{ijk}( b^2 ) b^2 B^{-1}_{ij} \partial^\mu \phi^{iI} \partial^\nu \phi^{jJ} \partial_\mu K^K_\nu  \label{lagns} \\ 
& {\cal L}_{Pol} = F(b-c\,b\,y^2) \label{lagpol}
\end{align}
As in \cite{Koichi}, there is a desire to study separately the spin by associating with microscopic scale while the scale of fluid remains for the orbital angular. However, such separation could lead to misinterpretation of the spin behavior in dissipative system. As the spin does not emerge from our theory by unitary expansion of quantum field theory, we need to "impose" further restrictions to the fluid \cite{polhydro1}. We define the polarization variable as $ y_{\mu \nu} \sim  u_\beta \partial^\beta \left( \sum_i \theta_i(\phi_I) \hat{T}_i  \right)$, where $\theta_i$, and $\hat{T}_i$ are the local phases and the generators, respectively. It is not accidental, at this point, the fact that the spin minimizes entropy when aligned with vorticity in local equilibrium $y_{\mu\nu} = \chi(b,\omega_{\mu\nu}\omega^{\mu\nu})\omega_{\mu\nu}, $ \cite{polhydro1}. Here $\omega_{\mu\nu} = \frac{1}{2} (\partial_\mu u^\nu - \partial_\nu u^\mu )$ is the kinetic vorticity.

Let us introduce a perturbation around the hydrostatic equilibrium state up to the second order as mentioned in Section \eqref{background}.
After some calculations, in the linear approximation, each process is defined by a Lagrangian. The linearized system then reads

\begin{equation}\label{alllag}
\underbrace{ \vphantom{\frac{}{}} \alpha_1 ( \ddddot \pi^I -  \partial_j^2 \ddot \pi^I+ \partial_I \partial_J\ddot\pi^J }_{Pol}) - \underbrace{  
 \alpha_2 \partial_I \partial_J \dot \pi^J}_{NS} - \underbrace{ \alpha_3 \partial_I \partial_J \dot \pi^J}_{Bulk} + \underbrace{ \ddot \pi^I - \alpha_4 \partial_I \partial_J \pi^J}_{Ideal} =0, 
\end{equation}
where the coefficients are $ \alpha_1 = 4 \chi^2(b_o^2,0), \ \alpha_2 = z(b_o^2) - 4 z^\prime (b_o^2), \ \alpha_3 =  2 h ( b^2_0 ) - h^\prime ( b^2_0 ), \ \alpha_4 = c^2_s = \frac{b^2_o F^{\prime\prime} (b_o,0) + \frac{b_o}{2} F^\prime(b_o,0)}{\frac{b_o}{2} F^\prime(b_o,0)} $. It is sometimes useful to study the system by splitting the perturbation into two orthogonal directions. Let's first analyze the dispersion relation for transverse perturbation. In this form, the propagation of perturbation is transverse to the sound waves

\begin{equation}\label{transeq}
\alpha_1 ( \omega_T^4 - k^2 \omega_T^2) + i \alpha_2   k^2 \omega_T - \omega_T^2 = 0,
\end{equation}
as shown in \cite{polhydro2,causality} these relations are non-causal, necessitating a relaxation time in a similar manner to Israel-Stewart equations.

\begin{figure}[h]\label{corrleo}
\includegraphics[width=8cm]{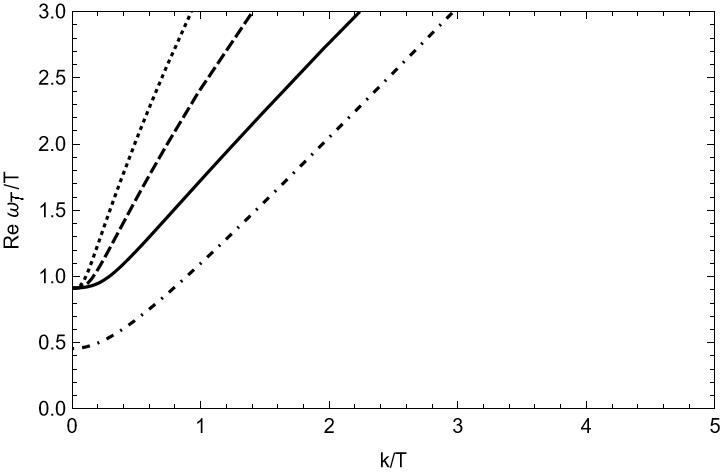}
\centering
\caption{ The dispersion relation for a polarizeable fluid with shear viscosity. black lines show Full, dotted, and dashed lines are $\eta = \{5,19.3,48\}$, respectively. Dot-dashed: the polarization without the shear contribution. For all lines, we use $\chi = 2.2$. \label{shearonly}}
\end{figure}
We see that the hydrodynamic limit increases when shear viscosity is added to the system, and the ratio $\omega/k$ that also display the internal energy dissipation within fluid. The gap in the hydrodynamic limit $\lim_{k \to 0} \omega (k) \neq 0$ just reinforce the fact that even the spin is treated as hydro-variables, it still lack a microscopic description or the fact that our approximation breaks down. We consider the so-called \textit{stability condition} in next section when we include the relaxation time for all dissipative processes. The Eq. \eqref{alllag} also gives the dispersion relation for longitudinal perturbation.

\begin{equation}\label{soundeq}
\alpha_1  \omega_L^4  + i ( \alpha_2  +  \alpha_3 ) k^2 \omega_L - \omega_L^2 + c_s^2 k^2 = 0.
\end{equation}

\begin{figure}[h]
\includegraphics[width=8cm]{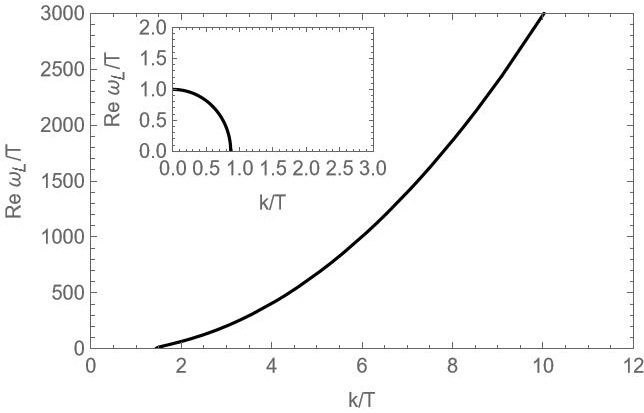}
\centering
\caption{Big panel shows the polarization with shear viscosity. Small panel shows the polarization without the shear contribution, as in \cite{polhydro2}\label{nonhydro}. For all lines, we use $\chi = 2.2$.}
\end{figure}

We are interested in analyzing the qualitative behavior of the dissipative processes of our polarizable fluid. As discussed in \cite{polhydro2}, there is a limited region $0 < k < \frac{\sqrt{F^\prime}}{8 c_s \chi}$ (small graphic) for which two of the four propagating modes are sound waves. In other case, they become nonpropagating modes, dependent of $k/T$. When we introduce the viscosity forces, the following result in the big graphic shows that the critical wavelength disappears, and the dispersion relation reproduces the interaction between polarization and viscosity. In other words, the dissipative forces changes the direction of spin in order to let the vortex-spin coupling aligned with the sound waves.

The results in this section are summarized in Fig. \eqref{shearonly} and \eqref{nonhydro}.  As can be seen in the Navier-Stokes limit the presence of polarization enhances the propagation of perturbations of energy-momentum, and hence the applicability of hydrodynamics.
But As Fig \eqref{shearonly} shows the relaxation time is still necessary to restore causality at larger $k$.  In the next sections we shall examine this.

\subsection{Including relaxation time}

We shall now combine the insights of \cite{causality} about what relaxational dynamics should look like with the techniques developed in \cite{lagorig}, including shear and bulk viscosity. Such formalism is well-known in the literature.

\begin{align}\label{all:action2}
 {\cal L} &= {\cal L}_{free} + {\cal L}_{IS-shear} + {\cal L}_{IS-Bulk} + {\cal L}_{IS-pol},  \\      
{\cal L}_{IS-bulk}  &=      \frac{\tau_\xi}{2}  \left(\Pi_- u^\alpha_+ \partial_\alpha \Pi_{ +} -  \Pi_+ u^\alpha_- \partial_\alpha \Pi_{ -}\right)   +   \frac{\Pi_{\pm}^2}{2}   +  h_{IJK}( b^2 ) K^{\mu I} K^{ \nu J} \partial_\mu K^K_\nu  , \label{eqsbulk2}   \\
{\cal L}_{IS-shear} &=  \frac{\tau_\eta}{2}  \left({\pi^{\mu\nu}_-} u^\alpha_+ \partial_\alpha {\pi_{\mu\nu}^{+}} - {\pi^{\mu\nu}_+} u^\alpha_- \partial_\alpha {\pi_{\mu\nu}^{-}} \right)   +   \frac{{\pi^{\mu\nu}_{\pm}}^2}{2}   + z_{IJK}( b^2 ) b^2 B^{-1}_{ij} \partial^\mu \phi^{iI} \partial^\nu \phi^{jJ} \partial_\mu K^K_\nu  , \label{eqsshear2}  \\
{\cal L}_{IS-pol} &=     \frac{\tau_\chi}{2}  ( \, Y^{\mu\nu}_{-} \, u^{\alpha}_{+} \partial_\alpha Y_{\mu\nu}^{+} - Y^{\mu\nu}_{+} u^{\alpha}_{-} \partial_\alpha Y_{\mu\nu}^{-} ) + \frac{Y^{\mu\nu 2}_{\pm}}{2} + F (b ( 1-c y^2 ))  . \label{eqspol2}
\end{align}
where $\{ \pi^{\mu\nu}_{\pm},\Pi_{\pm} , Y^{\mu\nu}_{\pm} \}$ are the new d.o.f. relaxes to the dissipative hydrodynamics-currents $\{\pi^{\mu\nu}_{\textit{Shear}},\Pi_{\textit{Bulk}} , y^{\mu\nu}_{\textit{Pol}} \}$  which can be expressed in terms of $\phi_I$ \cite{lagorig} and \cite{polhydro1} $y_{\mu \nu}$ (see Eqs \eqref{lagns} and \eqref{lagbulk} for the first two terms and Eq. \eqref{lagpol} for the third).

As shown in \cite{koide}, the causality structure of such theories depends on the dimensionless ration combining relaxation time, enthalpy and viscosity, e.g. 
$\tau_\eta(e+p)/\eta$.  In this paper, for simplicity, we therefore measure $\tau_{\eta,\zeta}$ in units of the 4th root of enthalpy, so $\tau_{\eta,\zeta}/(\eta,\zeta)$ is dimensionless.
\section{Dispersion relation of small perturbations \label{disp0}}
\subsection{Bulk viscosity dynamics}\label{section:bulk}

For the sake of simplicity, we start with the on-shell equations of motion for bulk current in IS theory. After applying the Euler Lagrange equation \eqref{eqsbulk2}, we obtain

\begin{equation}\label{bulk:eom}
\tau_{\xi}  u_\alpha \partial^\alpha \Pi  + \Pi    =  \Pi_{Bulk} = \Delta^{\mu \nu} \partial_\mu u_\nu = \partial_\gamma u^\gamma. 
\end{equation}
To evaluate the process consider the expansion of bulk viscosity, and neglecting the coefficient expansion. We get from the right hand side of the equation above the terms of the first and second order in $\pi$,

\begin{eqnarray}\label{bulk:exp}
   \tau_{\xi}  u_\alpha \partial^\alpha \Pi  + \Pi   & \simeq & 
   \Pi_{Bulk} (\mathcal{O} (\pi) ) + \Pi_{Bulk} (\mathcal{O} (\pi^2) ) \nonumber \\
  & \simeq &  \delta^\gamma_0 \left(  
 \dot \pi \partial_\gamma \dot \pi \right) + \delta^\gamma_I \left( - \partial_\gamma \dot \pi^I +  \partial_\gamma \dot \pi \cdot \partial\pi^I + \dot \pi \cdot  \partial \partial_\gamma \pi^I  \right) , 
\end{eqnarray}
By applying the Fourier transformation of the Eq. \eqref{ftdr} in the equation above, we obtain the dispersion relation for first and second order of the bulk viscosity    

\begin{eqnarray}\label{bulk:fourier}
\tilde{\Pi} (\pi) = \frac{1}{1 + i \omega_\xi \tau_\xi}  &\bigg\{& \underbrace{ - \omega k \pi_L   e^{ i\omega_{L} t - i{\bf k_L}\cdot {\bf x} } }_{ \sim \ \pi_L} - \underbrace{ i \omega^3 \pi_L^2  e^{i\omega_{L} t - i{\bf k_L}\cdot {\bf x}}}_{\sim \ \pi_L \pi_L} - \underbrace{ i \omega k^I k^J \pi^2_T  e^{i\omega_{T} t - i{\bf k_T}\cdot {\bf x}}}_{\sim \ \pi_T \pi_T } \nonumber \\
&&  - \underbrace{ i \omega k^I k^J \pi_T \pi_L e^{i(\omega_{L} + \omega_{T}) t - i ( {\bf k_T}\cdot {\bf x} + {\bf k_{L}}\cdot {\bf x} ) }}_{\sim \ \pi_T \pi_L } \bigg\}.
\end{eqnarray}    
The first and second term are the contribution for the sound perturbation and the third and fourth ones will be analyzed to study the dependence of polarization on bulk viscosity \cite{Schenke}. The fact that we have $\pi_T$ and $\pi_L$ does not imply they have different magnitude, but rather have different directions, which is not the same for NS contribution. We can finally find the Linearized version of Lagrangian \eqref{eqsbulk2}

\begin{align}\label{bulk:lag2}
{\cal L}_{Bulk} & \simeq - \bigg\{ h (b_0^2)  + b_0^2 h^\prime(b_0^2) ( [\partial\pi] +\frac{1}{2}[\partial\pi]^2 -\frac{1}{2}[\partial\pi\cdot\partial\pi] -\frac{1}{2} \dot\pi^2 )  + \frac{b_0^4}{2}  h^{\prime\prime}(b_0^2) [\partial\pi]^2    \bigg\}   \bigg\{ 1 + 3[\partial\pi] + \frac{9}{2} [\partial\pi]^2  \\ \nonumber 
&   - \frac{3}{2}  [\partial\pi\cdot \partial\pi] - \frac{3}{2}  \dot\pi^2 \bigg\} \bigg\{ \Pi(\mathcal{O} (\pi)) + \Pi(\mathcal{O} (\pi^2))  \bigg\}.
\end{align}
The dispersion relation including polarizable fluid is 

\begin{equation}\label{bulk:dp}
\frac{ \omega^2}{(1 + i \omega \tau_\chi)^2}  ( k_{\nu} k^\nu ({\bf k}\times {\boldsymbol \Omega^0})^I + k^I k^J ({\bf k}\times {\boldsymbol \Omega^0})^J) + \frac{i \omega }{(1 + i \omega \tau_\xi)}( A k^2 - B     \omega^2 ) - C \omega^2 + D k^2 = 0,
\end{equation}
where the coefficients are $ A^\prime = \frac{ 2 h( b^2_0 ) - h^\prime ( b^2_0 )}{4 c \, \chi^2(b_o^2,0)} $,  $ B^\prime = \frac{h(b_o^2)}{4 c \, \chi^2(b_o^2,0)} ,   C  =  \frac{F^\prime (b_o)}{4 c \, \chi^2(\alpha^3,0)} , D = \frac{1}{2} \frac{b_o^2 F^{\prime\prime} (b_o^2 )}{4 c \, \chi^2(b_o^2,0)} $
The analytical solution of $\omega$ is a little bit complicated with respect to the wave numbers then we will examine their limits forms.

\subsection{Evolution of a vortex in a fluid with bulk viscosity}

\subsubsection{Transverse modes}

The dispersion relations for $k \ll T$ is

\begin{equation}
\omega_T =\left\{ 
\begin{array}{c}
(2 + 6 A^\prime \tau_\xi + 18 A^\prime \tau_\chi) / \tau_\chi +  (A^\prime \pm k \tau_\chi)/6
\\ 
\frac{A^\prime(\sqrt{3}  + i)}{3 \tau_\xi - 9  \tau_\chi} \pm k \bigg( \frac{\sqrt{3} \tau_\chi}{3 \tau_\xi - 9 \tau_\chi} + \frac{3\tau_\chi}{2 + 6 \tau_\xi - 18 A^\prime \tau_\chi} \bigg) 
\end{array}%
\right. ,  \label{drr:bulk:small}
\end{equation}
We have four solutions whose propagating modes depend on $k$. In our effective field theory approach, the inclusion of polarization extinguishes the non-propagating modes. It is clear from the dispersion relation that the interaction between different dissipative process gives restrictions about their relaxations time. In principle, we should expect an isotropy of bulk viscosity. However, the spin current breaks the spherical symmetry in dynamics due to the fluid-mechanism of spin-vorticity coupling. As a consequence, the longitudinal and transverse directions of the fluid-matter suffer effects from different thermodynamics forces magnitude.

\begin{figure}[h]
  \begin{subfigure}{.45\textwidth}
  \centering
    \includegraphics[width=.8\linewidth]{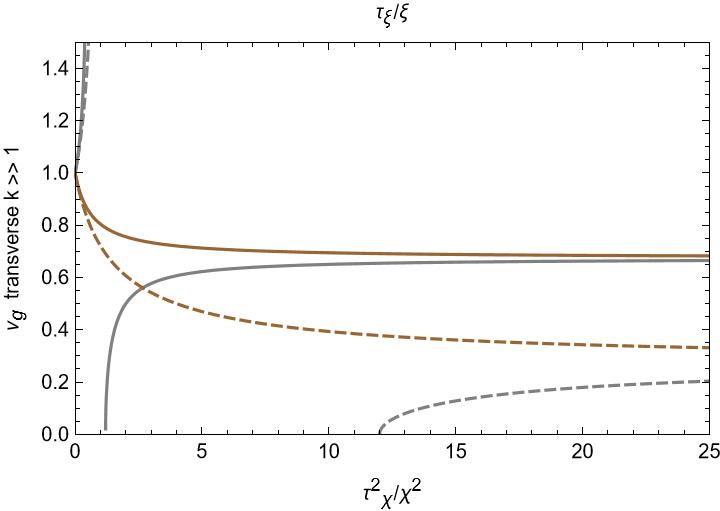}
  \end{subfigure}%
  \begin{subfigure}{.45\textwidth}
  \centering
    \includegraphics[width=.9\linewidth]{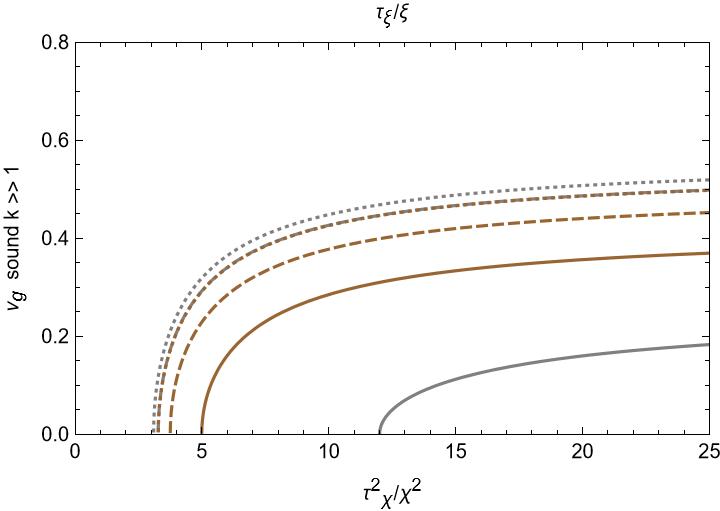}
  \end{subfigure}
   \caption{\textbf{Left:} The transverse modes of the asymptotic value of the group velocities. Brown: The full, dotted, and dashed lines are $\tau^2_\chi / \chi^2 = \{0.6, 1.2, 20.2\} $, respectively. Gray: The full, dotted, and dashed lines are $ \tau_\xi / \xi = \{1.2, 5.4, 57.2\}$, respectively. \textbf{Right:}  The longitudinal modes of the asymptotic value of the group velocities. Brown: The full, dotted, and dashed lines are $\tau^2_\chi / \chi^2 = \{7.52,15,35.2\} $, respectively. Gray: The full, dotted, and dashed lines are $ \tau_\xi / \xi = \{4,35.8, 90\}$, respectively.
    \label{translong1} }
\end{figure}

We can finally find the group velocity for asymptotic value of $k\rightarrow \infty$  (w.r.t. the temperature $T \sim b^{1/3}$ for dimensional analysis)

\begin{equation}\label{gp:bulk:trans}
\nu_g= \lim_{ k \gg T  } \frac{ d | \omega_T |}{d k} =  \frac{1}{\sqrt{ 1 +  A^\prime \frac{\tau^2_\chi}{\tau_\xi}  + C \tau^2_\chi } }  
\end{equation}
Our theoretical formulation \textit{at the level of Lagrangian} is well-suited for system that exhibits multiscale structure. It means how the microscale interactions interferes in the macroscopic scale behavior. For such a task, it requires to identify the symmetry breaking with dissipative processes, and the rest is automatic. The focus here is to clarify how the bulk viscosity influence the polarization, and consequently the macroscale structure. As shown in the left of Fig. \eqref{translong1}, the positive alignment between the bulk modes and polarization shifts $v_g$ away from the maximum value, while the opposite direction shifts the $v_g$ away from origin when we increase $\tau_\chi$. In both situations, $v_g$ reach a limit $\propto 1 - \tau_\xi / \xi$ when $\tau_\chi \gg \tau_\xi$ .

\subsubsection{Longitudinal modes}

The dispersion relations for $k \ll T$

\begin{equation}
\omega_L =\left\{ 
\begin{array}{c}
\pm  \frac{\sqrt{C - A^\prime k^2 \tau_\xi + k^4 \tau^2_\chi }}{(1 - k^2 (4A^\prime + B^\prime ) \tau_\chi  )} + \frac{i k^2 \tau_\chi }{2( C - \tau_\chi k^2 )}  \\ 
 \pm \bigg( kD + \frac{1}{2}\sqrt{C/(4 A^\prime + B^\prime ) + k^2  \tau_\xi }) +  i  \frac{\sqrt{8(4A^\prime + B^\prime ) - k^2  \tau_\xi}}{(C - k^2 (4A^\prime + B^\prime )\tau_\chi )}
\end{array}%
\right. ,  \label{drr:bulk:small}
\end{equation}
We have four solutions: All of them are propagating modes. The dispersion relations for $k \gg T$ is

\begin{equation}
\omega_L =\left\{ 
\begin{array}{c}
i/\tau_\xi \\
i/\tau_\chi + (\frac{1}{2} + i) \frac{(4A^\prime +B^\prime )}{4 D} +  \mathcal{O}(k^{-1})  \\ 
\pm k \sqrt{\frac{ D \tau^2_\chi + 1 + ( B^\prime + 8 A) \frac{\tau^2_\chi}{\tau_\xi} }{ C \tau^2_\chi - 1 - A^\prime \frac{\tau^2_\chi}{\tau_\xi} } } + i/(2(4A^\prime + B)\tau_\xi + \tau_\chi) + \mathcal{O}(k^{-1})
\end{array}%
\right. ,  \label{drr:bulk:large}
\end{equation}
For larger $k$, the imaginary part is always positive and we obtain a finite sound waves for all values of $k$. It should be stressed the spin contribution is proportional to $k^{-2}$, while bulk viscosity is  $\propto k^{-1}$. The  causality is guaranteed by the positiveness of the imaginary part.  It is easy to relate the vanish of stability in bulk Lagrangian when we pay attention to the double term $K^\mu K^\nu $, which let the action unbounded from below. Even though the $\partial_\mu K_\nu $ preserves homogeneity, isotropy, and parity symmetry, the break of $SO(3)$ by polarization leads to an anisotropy in the transient flow, as we can see in the dispersion relation \eqref{drr:bulk:small} and \eqref{drr:bulk:large}. The group velocity of sound waves is

\begin{equation}\label{gp:bulk:lg}
\nu_g= \lim_{ k \gg T  } \frac{ d | \omega_L |}{d k} =  \sqrt{\frac{ D \tau^2_\chi + 1 + ( B^\prime  + 8 A) \frac{\tau^2_\chi}{\tau_\xi} }{ C \tau^2_\chi - 1 - A \frac{\tau^2_\chi}{\tau_\xi} } },  
\end{equation}
the resulting causality constraint on relaxation time is

\begin{equation}\label{ineq:bulk}
\tau_\xi  \geq \frac{  \xi}{(1- c_s^2) (\epsilon + P) -  c \frac{  8  \chi^2}{\tau^2_\chi} }  .  
\end{equation}
Restoring the unpolarizeable medium by setting ($\chi \rightarrow 0$) \cite{Romatschke}. Our result suffers a slightly modification because of the spin diffusion. Here, the evolution in time of vorticity in transient flows dictates the accumulation of spin by the spin-vortex mechanism. Furthermore, such accumulation introduces an anisotropy turning the transport coefficient dependent of angular momentum. This is exactly what had to be expected. Thus, we have two sets of transport coefficients, one aligned $c>0$ and the other one anti-aligned $c<0$ with vortex. The inequality above tells us the direction of spin increase or decrease the relaxation time of bulk viscosity. 

Physically, for $c>0$ the presence of polarization lengthens the relaxation time for bulk viscosity w.r.t. the corresponding viscous force, modifying the fluctuation-dissipation relation.   This goes in the direction of the result of the previous section, in that the presence of polarization makes viscosity less apparent in the evolution of the system (a long relaxation time means viscous forces turn on after a longer time, allowing the perturbation to propagate).

\subsection{Shear viscosity\label{section:ns}}

In the precedent subsection, we apply the EFT techniques for the bulk viscosity. As the previous treatment can be applied here, we examine now the shear viscosity in a polarizable medium. Starting from the on-shell equations of motion 

\begin{equation}\label{ns:eom}
\tau_{\eta} u_\alpha \partial^\alpha \Pi^{\mu \nu} +  
\Pi^{\mu \nu} =  
\Delta^{\mu \nu \alpha \beta} \partial_\alpha u_\beta,  
\end{equation}
where $ \Delta^{\mu \nu \alpha \beta} = \frac{1}{2}( \Delta^{\mu\alpha} \Delta^{\nu\beta} + \Delta^{\mu\beta} \Delta^{\nu\alpha} - \frac{2}{3}
\Delta^{\mu\nu}\Delta^{\alpha\beta}) $, and $ \Delta^{\mu \alpha } \Delta^{ \nu \beta} \partial_\alpha u_\beta = \partial_\nu u_\mu - u_\nu u_\alpha \partial_\alpha u_\mu $. If we take into account just terms up the first and second order in $\pi$, we get from the right hand side of the equation above 

\begin{equation}\label{ns:exp}
\tau_{\eta} u_\alpha \partial^\alpha \Pi^{\mu \nu} +  
\Pi^{\mu \nu}   \simeq    \Pi_{Shear}^{\mu\nu} (\mathcal{O} (\pi)) + \Pi_{Shear}^{\mu\nu} (\mathcal{O} (\pi^2)).   
\end{equation}
The dispersion relation will take a slightly different resolution from the previous section where the direction will be left in the vectorial notation.

\begin{eqnarray}\label{ns:fourier}
\tilde{\Pi}^{\mu\nu}_{Shear} \mathcal{O}(\pi) & = & \frac{ \ \pi^I }{(1 + i \omega \tau_\eta)} \bigg\{ \underbrace{  -i(   \omega \delta^\nu_I  k^\mu +   \omega \delta^\mu_I  k^\nu )  e^{ i \omega t - i{\bf k}\cdot {\bf x}  } }_{\sim \ \vec{\pi}_1 \vec{\pi}_2 }  + \underbrace{ \frac{2}{3} ( g^{\mu\nu} - g^{\mu}_0 \, g^{\nu}_0 ) \omega k e^{ i\omega t - i{\bf k }\cdot {\bf x }  } }_{\sim \ \vec{\pi}_L \vec{\pi}_L } - \underbrace{ ( \delta^\mu_0  \delta^\nu_I + \delta^\nu_0  \delta^\mu_I ) \omega^2 e^{ i \omega t - i{\bf k}\cdot {\bf x}  } }_{\sim \ \vec{\pi} \vec{\pi} }      \bigg\} \nonumber \\
\tilde{\Pi}^{\mu\nu}_{Shear} \mathcal{O}(\pi^2) &=& \frac{\pi^I }{(1 + i \omega \tau_\eta)} \bigg\{ \underbrace{[  (- \delta^\nu_0 k^\mu \omega^2  + 2\delta^\nu_I k^\mu \omega k) +   ( -\delta^\mu_0 k^\nu \omega^2  + 2\delta^\mu_I k^\nu \omega k )] e^{ i \omega t - i{\bf k}\cdot {\bf x}  } }_{\sim \ \vec{\pi}_1 \vec{\pi}_2 } + i \frac{2}{3} ( g^{\mu\nu} - g^{\mu}_0 \, g^{\nu}_0 )  (\omega^3 + 2 k^2 \omega) \nonumber \\
&& \qquad \qquad + i \frac{2}{3} (  g^\mu_0 \, g^\nu_J + g^\mu_J \, g^\nu_0  ) \omega^2 k  + \underbrace{ i 2 (  \delta^\mu_0 \delta^\nu_0 + \delta^\mu_I \delta^\nu_J ) \omega^3 e^{ i\omega_L t - i{\bf k_L }\cdot {\bf x }  }   }_{\sim \ \pi_L \pi_L } - \underbrace{ i 3 ( \delta^\mu_0 \delta^\nu_I + \delta^\nu_0 \delta^\mu_I ) \omega^2 k e^{ i\omega_T t - i{\bf k_T }\cdot {\bf x }  }    }_{\sim \ \pi_T \pi_T }      \bigg\} .\nonumber \\
\end{eqnarray}    
We derive the terms of equation above in second and third derivative in order to get a linear approximation. This corresponds to a linear relation between the well-known NS term whose terms relevant for our calculation from \eqref{eqsshear2} are    
 
\begin{align}\label{bulk:lag2} 
{\cal L}_{NS} &=  z(b_o^2) \Pi^{\mu\nu} (\pi^2) + [b_o^2 z^\prime (b_o^2) + 3 z(b_o^2)] [\partial \pi] \Pi^{\mu\nu} (\pi). 
\end{align}
We can write the equation of motion of a NS polarizable fluid. The dispersion relation including polarizable fluid is 

\begin{align}\label{ns:dp}
&  \frac{ A }{(1 + i \omega \tau_\eta)} \bigg\{    (  2 \delta^\nu_I k^\mu \omega k - \delta^\nu_0 k^\mu \omega^2 ) +  ( 2 \delta^\mu_I k^\nu \omega k - \delta^\mu_0 k^\nu \omega^2 ) + i\frac{2}{3} ( g^{\mu\nu} - g^{\mu}_0 \, g^{\nu}_0 ) (\omega^3 + 2 k^2 \omega)  + i\frac{2}{3} (  g^\mu_0 \, g^\nu_J + g^\mu_J \, g^\nu_0  ) \omega^2 k + \nonumber \\ \nonumber \\ 
&  i2 (  \delta^\mu_0 \delta^\nu_0 + \delta^\mu_I \delta^\nu_J ) \omega^3   - i 3 ( \delta^\mu_0 \delta^\nu_I + \delta^\nu_0 \delta^\mu_I ) \omega^2 k  \bigg\}  +   \frac{i k B}{(1 + i \omega \tau_\eta)}  \bigg\{ i(   \omega \delta^\nu_I  k^\mu +   \omega \delta^\mu_I  k^\nu ) +  \frac{2}{3} ( g^{\mu\nu}  - g^{\mu}_0 \, g^{\nu}_0 )\omega k  -  ( \delta^\mu_0  \delta^\nu_I + \delta^\nu_0 \delta^\mu_I) \omega^2  \nonumber \\
&       \bigg\} -  \frac{\omega^2}{(1 + i \omega \tau_\chi )^2}  \bigg\{ k_{\nu} k^\nu ({\bf k}\times {\boldsymbol \Omega^0})^I + k^I k^J ({\bf k}\times {\boldsymbol \Omega^0})^J \bigg\}  + C \omega^2 - D k^2 = 0, \nonumber 
\end{align}
where the coefficients are $ A =  \frac{z(b_o^2)}{c \, \chi^2(b_o^2,0)}$ and $ B = \frac{ 3 z( b^2_0 ) - z^\prime ( b^2_0 )}{c \, \chi^2(b_o^2,0)}$.

\subsection{Evolution of a vortex in a fluid with shear viscosity}
 
The analytical solution of $\omega$ is a little bit complicated with respect to the wave numbers then we will examine their limits forms $(x \neq y = \{1,2,3\})$. 
 
\subsubsection{Transverse modes}
The dispersion relations for  $k\ll T$ is

\begin{equation}
\omega_T =\left\{ 
\begin{array}{c}
\frac{2 i  \tau_\chi}{3(C - 2 A \tau_\eta)} + \frac{3C \pm  k^2 (1 - 2 A \tau_\eta)}{(1 + k^2 \tau^2_\chi)} 
\\ 
 \frac{i}{8 \tau_\chi} + \frac{2A + 9 C \tau_\chi \pm k^2 (\tau_\chi + A \tau_\chi \tau_\eta)  }{(C + k^2 + 2 A k^2 \tau_\eta)}
\end{array}%
\right. ,  \label{drr:ns:small}
\end{equation}
The dispersion relation presents a shift in the hydrodynamic limit. The linear relation $\omega \propto k$ expected for weak dissipative system is no longer expressed in this transverse dispersion relation. Even though the spin-vorticiy coupling introduces terms of $\partial^4_t$ in the Lagrangian, the influence of $\tau_\chi$ is dominant for short wavelength. It theoretically suggest a multiscale physical framework for polarizable fluid. The positiveness of imaginary part is guaranteed for all wavelength. The transverse group velocity is

\begin{equation}\label{gp:ns:trans}
\nu^{(x,y)}_g= \lim_{ k \gg T  } \frac{ d | \omega_T |}{d k} =  \sqrt{ \frac{1 + 2A \frac{\tau^2_\chi}{\tau_\eta}  (\delta^x_I \delta^y_J + \delta^x_J \delta^y_I ) }{ 1 + C \tau^2_\chi + \frac{A \tau^2_\chi}{\tau_\eta} (\delta^x_I \delta^y_J + \delta^x_J \delta^y_I )  } },    
\end{equation}
where $x \neq y = \{1,2,3\}$. As shown in the left of Fig. \eqref{translong2}, even though the group velocity is acausal in some wavelength regime, the causality condition determines the  asymptotic value of group velocity must be less than one. The group velocity shows more sensitivity to influence of polarization than shear viscosity. 

\subsubsection{Longitudinal modes}

\begin{figure}[h]
     \centering
     \begin{subfigure}{0.45\textwidth}
         \centering         \includegraphics[width=\linewidth]{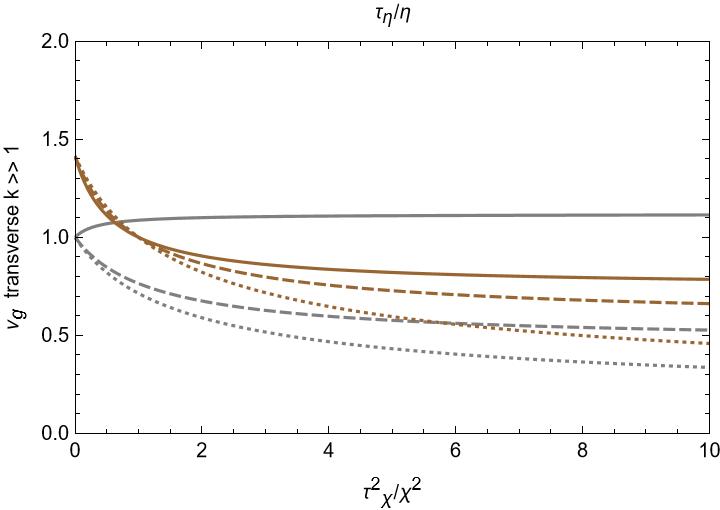}
     \end{subfigure}
     \begin{subfigure}{0.5\textwidth}
         \centering         \includegraphics[width=\linewidth]{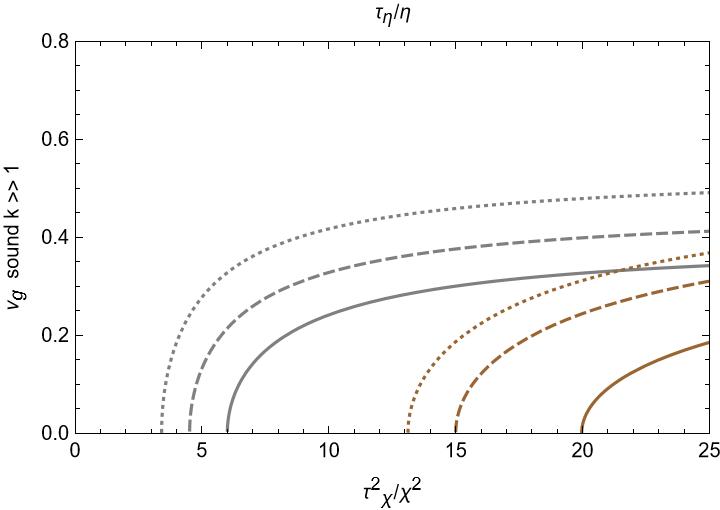}
     \end{subfigure}
       \caption{\textbf{Left:} The transverse modes of the asymptotic value of the group velocities. Brown: The full, dotted, and dashed lines are $\tau^2_\chi / \chi^2 = \{0.8, 2, 25.4\} $, respectively. Gray: The full, dotted, and dashed lines are $\tau_\eta / \eta = \{0.6, 8.4, 78\}$, respectively. \textbf{Right:}        The longitudinal modes of the asymptotic value of the group velocities. Brown: The full, dotted, and dashed lines are $\tau^2_\chi / \chi^2 = \{7.52, 15, 35.2\} $, respectively. Gray: The full, dotted, and dashed lines are $\tau_\eta / \eta = \{24,35.8,90\}$, respectively.
         \label{translong2}}
\end{figure}

The dispersion relations for $k \ll T$

\begin{equation}
\omega_L =\left\{ 
\begin{array}{c}
\frac{4 (A-C + B k^2 \tau_\eta )}{4 A (A - i D) \tau_\eta - 9 k^2 (B - 2i A \tau_\eta)} \pm k \frac{(3 A \tau_\eta (B - 2i A \tau_\eta))}{(2A - D)\tau^2_\chi} \\ 
\pm  k \bigg( D + \frac{ \sqrt{3(A - i C)(2A + i D)^3}}{3A}  \bigg)
\end{array}%
\right. ,  \label{drr:ns:small}
\end{equation}
The dispersion relations for $k \gg T$ is
\begin{equation}
\omega_L =\left\{ 
\begin{array}{c}
\frac{i8 D }{(2 A +\tau_\chi)}
\\
\frac{i B}{4 (D + 2 A \tau_\eta )}  \\ 
\pm  k \sqrt{\frac{D + 1/ \tau^2_\chi + (B+2A)/ \tau_\eta   }{ C  - 1/ \tau^2_\chi - A / \tau_\eta }} + \mathcal{O}(k^{-1})
\end{array}%
\right. ,  \label{:ns:small}
\end{equation}
For $k \gg T$, after $t \gg \{ \tau_\eta, \tau_\chi \}$, the system recovers his isotropy configuration. The damping transport coefficient of sound waves is positive in Eqs. \eqref{drr:ns:small} and \eqref{:ns:small}. When $\tau_\chi \gg \tau_\eta $, we recover the group velocity in \cite{causality}. We remark that \eqref{:ns:small} shows the maximum velocity of the plane-wave perturbation and the frequency damps exponentially by a constant of imaginary part. The group velocity of sound waves is 

\begin{equation}\label{gp:ns:lg}
\nu^{(x,y)}_g= \lim_{ k \gg T  } \frac{ d | \omega_T |}{d k} =  \sqrt{ \frac{D \tau^2_\chi + 1 + (B+2A) \frac{\tau^2_\chi}{\tau_\eta}  (\delta^x_I \delta^y_J + \delta^x_J \delta^y_I ) }{ C \tau^2_\chi - 1 - \frac{A \tau^2_\chi}{\tau_\eta} (\delta^x_I \delta^y_J + \delta^x_J \delta^y_I )  } }.  
\end{equation}
As shown in the right of Fig. \eqref{translong2}, we concern how the spin-vorticiy coupling interacts with the viscosity forces from the environment of the transient flows. As the formation of vortex is gapless for an unpolarized fluid \cite{nicolis}, the dissipationless polarizable fluid generates a ultraviolet cutoff in momentum space $\propto \chi^{-2} >0$, in this context, roughly speaking, the new "vacuum state", invariant under $SO(2)$ and symmetries in Eq. \eqref{first}, establishes a new gap and spin must be parallel to the vorticity \cite{causality}. The mass matrix $\delta^{ii}\chi^{-2}$, following the Kubo formulae $\chi^2   \propto \lim_{ \omega \rightarrow 0} \frac{1}{\omega} \int dt d^3 x \ave{ \left[ y_{ii}(x,t), y_{ii}(0) \right]} e^{  i(k x- \omega t) } $, is responsible for creating this gap in the vortex formation. Hence, the effective polarization is $\chi^{\prime -2} \rightarrow \chi^{-2} + \delta \chi^{-2}$, where $\chi^{\prime 2}$ is the direct dissipative mass due to polarization, and $\delta \chi^2$ is the correction due to the viscous tensor.  Hence, one seees the interaction in \cite{shear1,shear2} at the level of the dispersion relation, as expected in \cite{lin2}

\begin{equation}\label{ineq:ns}
\tau_\eta  \geq \frac{ 4 \eta}{(1- c_s^2) (\epsilon + P) -  c \frac{  8  \chi^2}{\tau^2_\chi} } .
\end{equation}
One again, we see the dynamics of Eq. \ref{ineq:bulk} where for $c>0$ a perturbation propagates for longer before being quenched, looking ''more hydrodynamical'', because viscous forces are delayed by the polarization.

Note the positiveness of $\chi$ only remains valid when we analyze the restrictions of entropy current for a dissipationless polarizeable fluid \cite{causality}. However, after including gradients forces of first order by effective action developed in \cite{lagorig}, $\eta$ affects the positiveness of $\chi$. It occurs because the transient flows generates a gradient of vortex accumulation. This variation in the spatial and orientational distribution of vortex produces a spin current due to spin-vorticiy coupling. We can characterize the generation of this spin current by two different process $i)$ the change of spin orientation based on pressure sound waves from transient flows $ii)$ spatial gradient of spin concentration between the fluid layers. Since the dynamics of our system is dictated by Lagrangians, this physical process comes at cost of having a broken symmetry from shear viscosity \cite{lagorig}. This geometric transformation of dynamics happens when we introduce the $B_{IJ}^{-1}$ to Eq. \eqref{eqsshear2}.  The interplay between $B_{IJ}$ and polarization can be thought of as symmetry breaking, due to polarization density being equivalent to a vacuum $B_{IJ}$ expecation value, as shown in \cite{lin1,lin2}

\section{Dispersion relations with polarization, bulk and shear viscosity\label{disp1}}

\textcolor{black}{In the preceding sections, we carried out our analysis to study the linearized dynamics of each dissipative process separately in a polarizable fluid, using methods described in \cite{causality}.   We are now ready to combine eqs \Cref{eqsbulk2,eqsshear2,eqspol2} to obtain a complete linearized dispersion relation}

\begin{figure}[h]
     \centering
     \begin{subfigure}{0.45\textwidth}
         \centering         \includegraphics[width=\linewidth]{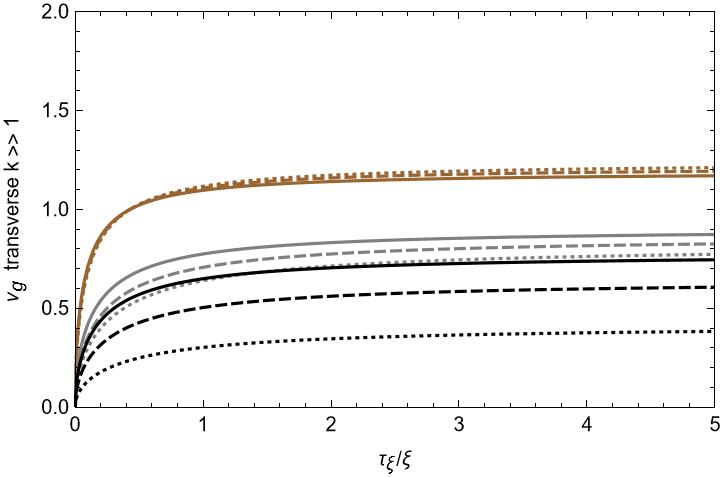}
     \end{subfigure}
     \begin{subfigure}{0.5\textwidth}
         \centering         \includegraphics[width=\linewidth]{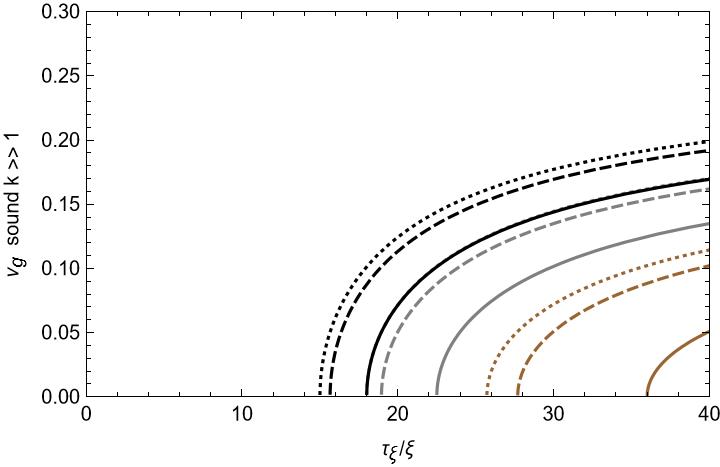}
     \end{subfigure}
           \caption{\textbf{Left:} The transverse modes of the asymptotic value of the group velocities. The full, dotted, and dashed lines are $\tau^2_\chi / \chi^2 = \{0.8,2,25.4\} $, respectively. Brown, Gray, and Black are $\tau_\eta / \eta = \{ 0.3, 2, 13 \} $, respectively. \textbf{Right:} The longitudinal modes of the asymptotic value of the group velocities. The full, dotted, and dashed lines are $\tau^2_\chi / \chi^2 = \{30, 40, 45\} $, respectively. Brown, Gray, and Black are $\tau_\eta / \eta = \{ 80, 120,180 \} $, respectively. 
         \label{translong3}}
\end{figure}

\subsection{Evolution of a vortex in a fluid with shear and bulk viscosity}

which results in a group velocity

\begin{equation}\label{gp:ns:bulk:trans}
\nu_g= \lim_{ k \gg T  } \frac{ d | \omega_L |}{d k} =  \sqrt{\frac{1 + A^\prime  \frac{\tau^2_\chi}{\tau_\eta} (\delta^x_I \delta^y_J + \delta^x_J \delta^y_I)}{ 1 + C \tau^2_\chi +   A \frac{\tau^2_\chi}{\tau_\xi} + A^\prime \frac{\tau^2_\chi}{\tau_\eta} (\delta^x_I \delta^y_J + \delta^x_J \delta^y_I)  } } 
\end{equation}
We can finally find the group velocity for asymptotic value of $k$ correspond to

\begin{equation}\label{gp:ns:bulk:lg}
\nu^{(x,y)}_g = \lim_{ k \gg T  } \frac{ d | \omega_T |}{d k} =  \sqrt{ \frac{D \tau^2_\chi + 1 + (B+2A) \frac{\tau^2_\chi}{\tau_\eta}   + ( B^\prime + 8 A^\prime ) \frac{\tau^2_\chi}{\tau_\xi} }{ C \tau^2_\chi - 1 - A \frac{\tau^2_\chi}{\tau_\xi} - A^\prime \frac{ \tau^2_\chi}{\tau_\eta}} }.
\end{equation}
These group velocities are plotted w.r.t. bulk and shear timescales in Fig. \ref{translong3}

\begin{figure}[h]
  \begin{subfigure}{.45\textwidth}
  \centering
    \includegraphics[width=.9\linewidth]{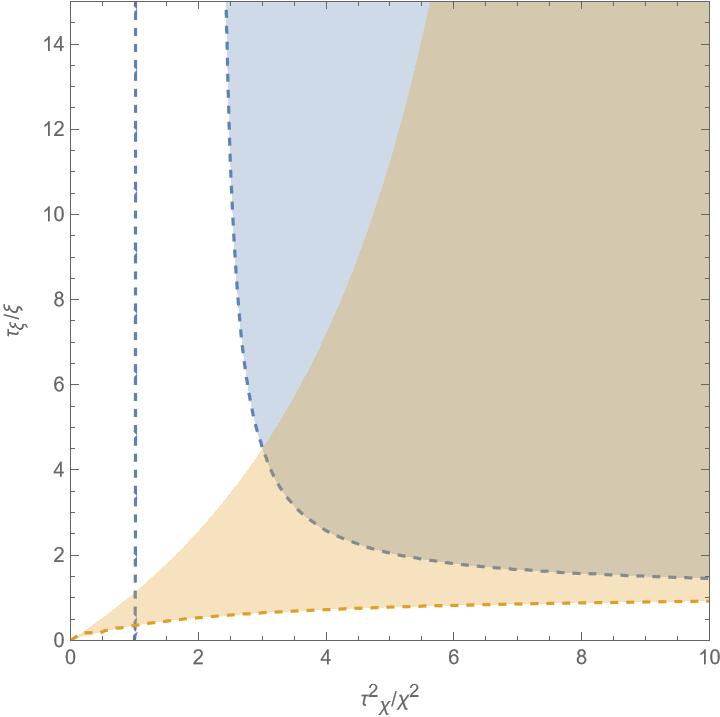}
  \end{subfigure}
  \begin{subfigure}{.45\textwidth}
  \centering
    \includegraphics[width=.9\linewidth]{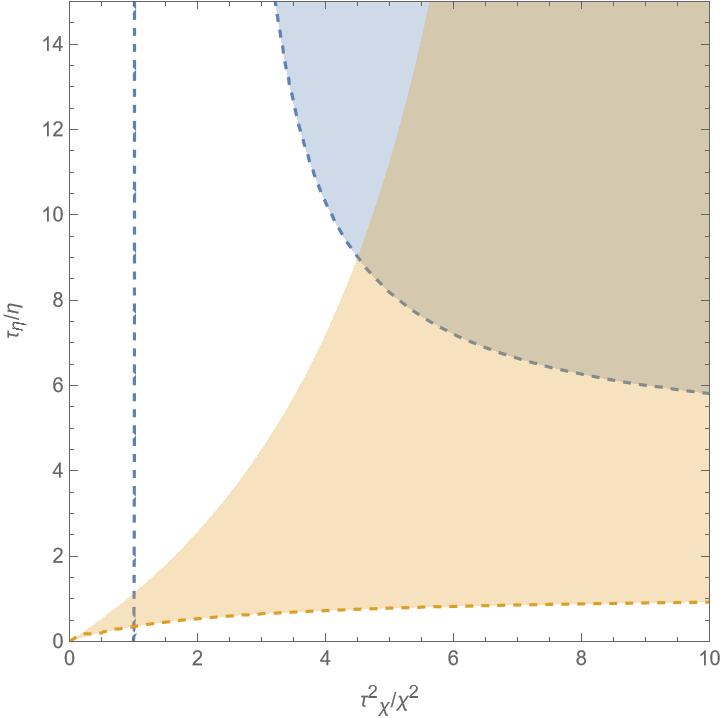}
  \end{subfigure}
  \caption{ Bulk (\textbf{Left:}) and  Shear (\textbf{Right:}) causally allowed regions for spins parallel (lighter) and antiparallel (darker) spin w.r.t. vorticity. A hydrodynamic evolution independent of initial conditions occurs at the intersection of all 4 regions}
  \label{colorplot}
\end{figure}

Causality then gives a constraint from causality mixing $\tau_\eta$ and $\tau_\xi$
 
\begin{equation}\label{ineq:nsbulk}
\tau_\eta  \tau_\xi  \geq \frac{\eta \tau_\xi + \xi \tau_\eta}{(1 - c_s^2 )(\epsilon + P)  - c \frac{8\chi^2}{\tau^2_\chi} }.
\end{equation}
 As in the previous sections,  this constraint illustrates why the presence of polarization lowers the effective viscosity.  As susceptibility becomes non-zero the time-scales of shear and bulk viscosity become longer,in natural units, w.r.t. the effective viscosity, 
 hence in a sense polarization prevents viscosity from being turned on.  Fig. \ref{colorplot} shows the effect of these constraints.  Because the constraint is multiplicative on the RHS and additive on the LHS, the largest viscosity (between shear and bulk) tends to be quenched more, enhancing even more apparent fluidity of a highly viscous but polarizeable medium.

\section{Discussion \label{disc}}
Let us try to draw a few qualitative lessons from the calculations of the previous two sections. First of all, we note that the polarizationless limit in a viscous fluid is not a trivial one. In order to eliminate the polarization of our system, we need to take $\chi \rightarrow 0$ faster than $\tau_\chi \rightarrow 0$.
However  Fig. \eqref{nonhydro} shows the applicability of hydrodynamics, defined by the presence of a real component of $w$ for a given $k$, in the case where just polarization is present w.r.t. the case of no polarization examined in \cite{polhydro2}.   The decrease of $w$ to zero signals when the ``sound-wave`` becomes a gapped non-propagating ``non-hydrodynamic mode``.  The result is surprising,and confirms that the interplay of the viscous scale and the hydrodynamic scale is highly non-trivial (as already seen in \cite{lin2}: adding shear viscosity on top of polarization {\em increases} the applicability of hydrodynamics, in the sense that perturbations beyond the critical frequency associated with the disappearance of the hydrodynamic mode in \cite{polhydro2} can propagate. This effect might have deep phenomenological consequences, as it might lead to an effectively low viscosity w.r.t. the expectation from applying the Kubo formula to calculate shear propagation, without polarization input.  

This conclusion is reinforced by looking at the asymptotic value of the group velocity. The left panels of figures \eqref{translong1} \eqref{translong2} and \eqref{translong3} show the asymptotic ($k\rightarrow \infty$ the transverse modes, and the right panels  show the corresponding longitudinal modes depend on the relaxation times for polarization, shear and bulk.  It can be seen that a lagrangian hydrodynamics with polarization can be UV complete, i.e. causal for arbitrarily high $k$ for certain values of the ``infrared`` fluctuation dissipation relations, namely long enough relaxation times w.r.t. the corresponding viscous forces.   The exact values where this happens however depends non-trivially on combinations of $\tau_{\chi,\eta,\xi}$.

\begin{figure}[h]
\centering 
\includegraphics[width=9cm]{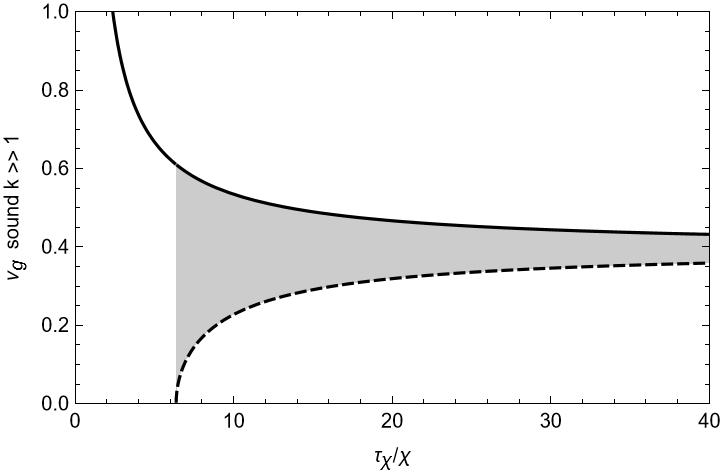}
\caption{The region of $\tau_\chi/\chi$ allowed by causality. The parameters are 12 8  \label{chiconfidence}}
\end{figure}

Once all dependence on viscosity is projected out, the causally allowed region depending on polarization is shown in Fig. \eqref{chiconfidence}.  As can be seen, the presence of viscous corrections give upper {\em and} lower limits to the group velocity.  Just like polarization decreases the effective viscosity, the presence of polarization imposes upper and lower constraints on the propagation of hydrodynamic perturbations.  
Interpreting the lower limit of the group velocity is at first sight problematic: What ''tells a slow perturbation that it can not exist''?   However, if one considers hydrodynamics as an effective theory of statistical mechanics in every coarse-graining cell, then the symmetry considerations in \cite{gauge,sampaio} can help clarify the situation: Fig \eqref{chiconfidence} says that slow perturbations can not evolve causally.   THis means that rather than dynamical propagating degrees of freedom these perturbations give rise to fluctuations that evolve stochastically.   The full dynamics of such small fluctuations will be given by a stochastically fluctuating theory along the lines of \cite{sampaio} left to be investigated in a forthcoming work, but it is already known that the interplay of the gradient \cite{groz} and stochastic \cite{farid} structure of non-equilibrium hydrodynamics is quite involved, in light with the considerations of this work.

As a related point, Eqs  \eqref{ineq:bulk},\eqref{ineq:ns} and \eqref{ineq:nsbulk} are potentially very important for defining the full analytical structure of the local equilibrium partition function: As we know, $\chi$ and $\tau_\chi$,as well as viscosity and heat capacities (connected to $\zeta$ and $\tau_\eta$) are connected by a fluctuation-dissipation relation \cite{kadanoff} which was investigated in \cite{lin2}.
This relation is ''galilean'', in that space and time are separated, while the equations above are Lorentzian, in that they keep track of causality in all frames of reference \cite{koide}.   As shown in \cite{sampaio} combining such relations with local Lorentz invariance requires in additional symmetries.    The results here, go in a similar direction, for they show that in a dissipative theory with spin causality in a Lorentz invariant sense imposes necessarily relations linking $\mathrm{Im}G_\chi/\mathrm{Re}_{\chi}$ with  $\mathrm{Im}G_{\Pi}/\mathrm{Re}_{\Pi}$ (where $G_X$ is the two-point function responsible for fluctuation and dissipation) $\Pi$ are the shear and bulk viscous tensor).   
This conclusion is further clarified by looking at Fig. \eqref{colorplot}.  The physical region is the one where causality is maintained for arbitrary $k$ independently of the initial spin orientation w.r.t. vorticity, i.e. the intersection of the shaded areas.   This plot therefore shows that causality imposes limits on how big or small can $\tau_\chi/\chi$ be w.r.t. the corresponding dimensionless ratio of the relaxation time to the shear and bulk tensor.
This strengthens our earlier causality argument \cite{causality} and motivates combining the sort of causality analysis done here with an analysis of the analytical structure of the theory \cite{groz}, including the role of fluctuations \cite{farid}.  This will be left to an upcoming work.

These results have an important phenomenological dimension: As \cite{lin2} demonstrated, spin-shear coupling of the type argued for in \cite{shear1,shear2} occurs when the spin relaxation timescale is parametrically longer than the viscosity one.    As can be seen from \eqref{chiconfidence} general causality considerations make this nearly inevitable, providing evidence that this process is indeed present in all systems where hydrodynamics applies.

 These results also add some clarity on whether the spin-shear coupling term is dissipative or not, something on which different approaches \cite{shear1,shear2} give different results.  As shown in \cite{lin2}, the shear-spin correction is determined by transport coefficients determined by shear viscosity.   But as As Fig. \eqref{nonhydro} shows , adding bulk and shear viscosity correct a dispersion relation that as we know from \cite{polhydro2,causality} is already dissipative.  
 Without an explicit propagation causality analysis, the dissipative nature of this transport coefficient can be misconstrued.
 This situation is actually well-known in non-relativistic hydrodynamics, famous examples being Darcy's law \cite{darcy} and reversible fluid mixing \cite{revfluid}: When dissipative dynamics dominates over non-dissipative, but there are two dissipative forces (respectively viscosity and heat conduction for \cite{darcy}, and viscosity and molecular diffusion in \cite{revfluid}), dynamics between them is paradoxically non-dissipative, something explicitly demonstrated in \cite{revfluid}.
 In our case, we have dissipation and spin relaxation, both dissipative.   In the limit that local dynamics is dominated by them interacting, we get an effective reversible dynamics.
 
While the numerical implementation of the equations developed here is some way away, some phenomenological lessons are in order: The lower effective viscosity of polarizeable systems suggests that polarizability is correlated with enhancement of observables sensitive to low-viscosity fluids, such as $v_n$.  It will be interesting to make this correlation experimentally, by finding classes of events with lower polarizability and seeing if viscosity in such classes of events is higher.  Since so far both polarizability and low-viscosity fluidity were found everywhere (the quenching of global polarization at higher energies being explained by higher transparency, and local polarization being present at these energies) we await lower energy runs for developments on this front.

On the theoretical side, while comparing Lagrangian to Eulerian type approaches term by term is generally non-trivial, the Zubarev statistical operator can provide such a bridge \cite{sampaio}.  
Recently, spin alignment has been calculated in the Zubarev statistical operator approach \cite{zubspin}.  The results, particularly the interplay between spin alignment and shear tensor, are compatible with ours.  It would be interesting to see whether the Zubarev approach can be matched to the lagrangian approach order by order (as shown in \cite{sampaio} the Lagrangian approach can indeed be seen as a limit of the Zubarev approach but fluctuation and gradient expansion generally fails to commute).  

Furthermore since, as argued in \cite{sampaio} causality violation might, rather than a symptom of a bad theory, be allowed provided we are dealing with fluctuating quantities and no-signaling backward in time happens, an analysis of the fully fluctuating theory could modify the results of this paper.  It is not clear whether spin and angular momentum are separately observable as components of the energy-momentum tensor, so part of the non-causal region might reveal itself to be physically irrelevant.  We find this unlikely as the results in this work, being Lagrangian, are pseudo-gauge invariant, but this needs to be explicitly checked.

Finally, it is well known that linear stability criteria might fail when higher order perturbative corrections \cite{fogaca} are included, and even then non-perturbative stability criteria could be quite different from perturbative ones \cite{disconzirev}.  Hence, resummed loop corrections to the dispersion relation, calculated wih the techniques developed in \cite{lin1,lin2} need to be examined with criteria beyond linear stability \cite{disconzirev}.  This is left to a follow-up work.

In conclusion, we have calculated the dispersion relation of small perturbations of a fluid characterized by vortical susceptibility, shear and bulk viscosity and respective relaxation times. We have also analyzed its causality structure.   The results are non-trivial in that they suggest that the timescales for each of these processes interplay in non-trivial ways.   In particular, polarization can decrease effective shear and bulk dissipation and causality imposes constraints on the relations between shear,bulk and polarization timescales.  The group velocity of perturbations is also bounded below as well as above.

These results could have phenomenological applications, particularly in the  low viscosity combined with non-negligible vortical susceptibility inferred from experiment, as well as the role of shear-polarization coupling.   We hope therefore to incorporate them in numerical models used to model heavy ion experimental data.

\textit{Acknowledgments}  G.T. acknowledges support from Bolsa de produtividade CNPQ 305731/2023-8, Bolsa de
pesquisa FAPESP 2023/06278-2. D.M.C thanks the JINR for full support.


\end{document}